\documentclass[aps,prl,twocolumn,showpacs,preprintnumbers]{revtex4}
\def\beq{\begin{equation}}
\def\eeq{\end{equation}}
\def\bea{\begin{eqnarray}}
\def\eea{\end{eqnarray}}
\def\nn{\nonumber}
\def\sss{\scriptscriptstyle}
\def\bd{B_d^0}
\def\bdbar{{\overline{B_d^0}}}
\def\bs{B_s^0}

\def\barp{{\raise.35ex\hbox
{${\sss (}$}}---{\raise.35ex\hbox{${\sss )}$}}}
\def\bdbarp{\hbox{$B_d$\kern-1.4em\raise1.4ex\hbox{\barp}}}
\def\bsbarp{\hbox{$B_s$\kern-1.4em\raise1.4ex\hbox{\barp}}}
\def\ks{K_{\sss S}}

\def\roughly#1{\mathrel{\raise.3ex\hbox
{$#1$\kern-.75em\lower1ex\hbox{$\sim$}}}}
\def\lsim{\roughly<}

\def\pewc{P_{\sss EW}^{\sss C}}
\def\pewcp{P_{\sss EW}^{\prime \sss C}}
\newread\epsffilein 
\newif\ifepsffileok 
\newif\ifepsfbbfound 
\newif\ifepsfverbose 
\newdimen\epsfxsize 
\newdimen\epsfysize 
\newdimen\epsftsize 
\newdimen\epsfrsize 
\newdimen\epsftmp 
\newdimen\pspoints 
\def\epsfbox#1{\global\def\epsfllx{72}\global\def\epsflly{72}%
 \global\def\epsfurx{540}\global\def\epsfury{720}%
 \def\lbracket{[}\def\testit{#1}\ifx\testit\lbracket
 \let\next=\epsfgetlitbb\else\let\next=\epsfnormal\fi\next{#1}}%
\def\epsfgetlitbb#1#2 #3 #4 #5]#6{\epsfgrab #2 #3 #4 #5 .\\%
 \epsfsetgraph{#6}}%
\def\epsfnormal#1{\epsfgetbb{#1}\epsfsetgraph{#1}}%
\def\epsfgetbb#1{%
\openin\epsffilein=#1
\ifeof\epsffilein\errmessage{I couldn't open #1, will ignore it}\else
 {\epsffileoktrue \chardef\other=12
 \def\do##1{\catcode`##1=\other}\dospecials \catcode`\ =10
 \loop
 \read\epsffilein to \epsffileline
 \ifeof\epsffilein\epsffileokfalse\else
 \expandafter\epsfaux\epsffileline:. \\%
 \fi
 \ifepsffileok\repeat
 \ifepsfbbfound\else
 \ifepsfverbose\message{No bounding box comment in #1; using defaults}\fi\fi
 }\closein\epsffilein\fi}%
\def\epsfclipstring{}
\def\epsfsetgraph#1{%
 \epsfrsize=\epsfury\pspoints
 \advance\epsfrsize by-\epsflly\pspoints
 \epsftsize=\epsfurx\pspoints
 \advance\epsftsize by-\epsfllx\pspoints
 \epsfxsize\epsfsize\epsftsize\epsfrsize
 \ifnum\epsfxsize=0 \ifnum\epsfysize=0
 \epsfxsize=\epsftsize \epsfysize=\epsfrsize
 \epsfrsize=0pt
 \else\epsftmp=\epsftsize \divide\epsftmp\epsfrsize
 \epsfxsize=\epsfysize \multiply\epsfxsize\epsftmp
 \multiply\epsftmp\epsfrsize \advance\epsftsize-\epsftmp
 \epsftmp=\epsfysize
 \loop \advance\epsftsize\epsftsize \divide\epsftmp 2
 \ifnum\epsftmp>0
 \ifnum\epsftsize<\epsfrsize\else
 \advance\epsftsize-\epsfrsize \advance\epsfxsize\epsftmp \fi
 \repeat
 \epsfrsize=0pt
 \fi
 \else \ifnum\epsfysize=0
 \epsftmp=\epsfrsize \divide\epsftmp\epsftsize
 \epsfysize=\epsfxsize \multiply\epsfysize\epsftmp
 \multiply\epsftmp\epsftsize \advance\epsfrsize-\epsftmp
 \epsftmp=\epsfxsize
 \loop \advance\epsfrsize\epsfrsize \divide\epsftmp 2
 \ifnum\epsftmp>0
 \ifnum\epsfrsize<\epsftsize\else
 \advance\epsfrsize-\epsftsize \advance\epsfysize\epsftmp \fi
 \repeat
 \epsfrsize=0pt
 \else
 \epsfrsize=\epsfysize
 \fi
 \fi
 \ifepsfverbose\message{#1: width=\the\epsfxsize, height=\the\epsfysize}\fi
 \epsftmp=10\epsfxsize \divide\epsftmp\pspoints
 \vbox to\epsfysize{\vfil\hbox to\epsfxsize{%
 \ifnum\epsfrsize=0\relax
 \includegraphics{#1}%
 \else
 \epsfrsize=10\epsfysize \divide\epsfrsize\pspoints
 \includegraphics{#1}%
 \fi
 \hfil}}%
\global\epsfxsize=0pt\global\epsfysize=0pt}%
 {\catcode`\%=12 \global\let\epsfpercent=
\long\def\epsfaux#1#2:#3\\{\ifx#1\epsfpercent
 \def\testit{#2}\ifx\testit\epsfbblit
 \epsfgrab #3 . . . \\%
 \epsffileokfalse
 \global\epsfbbfoundtrue
 \fi\else\ifx#1\par\else\epsffileokfalse\fi\fi}%
\def\epsfempty{}%
\def\epsfgrab #1 #2 #3 #4 #5\\{%
\global\def\epsfllx{#1}\ifx\epsfllx\epsfempty
 \epsfgrab #2 #3 #4 #5 .\\\else
 \global\def\epsflly{#2}%
 \global\def\epsfurx{#3}\global\def\epsfury{#4}\fi}%
\def\epsfsize#1#2{\epsfxsize}
\begin{document}

\preprint{UdeM-GPP-TH-04-120}
\preprint{McGill 07/04}
\preprint{UAB-FT-558}

\title{Testing the Standard Model with {\boldmath $\bs\to K^+ K^-$} Decays}
\author{David London}
\affiliation{Physics Department, McGill University, 3600 University
St., Montr\'eal QC, Canada H3A 2T8; \\ Laboratoire Ren\'e
J.-A. L\'evesque, Universit\'e de Montr\'eal, C.P. 6128,
succ.~centre-ville, Montr\'eal, QC, Canada H3C 3J7}
\author{Joaquim Matias}
\affiliation{IFAE, Universitat Aut\`onoma de Barcelona, 08193
Bellaterra, Barcelona, Spain}
\date{\today}
\begin{abstract}
In the limit of flavor SU(3) symmetry, the amplitudes for
$\bd\to\pi^+\pi^-$ and $\bs\to K^+ K^-$ are related to one another.
Taking the values of the weak phases $\beta$ and $\gamma$ from
independent measurements, the ratio $R_d^s \equiv BR(\bs \to K^+
K^-)/BR(\bd \to \pi^+ \pi^-)$ and the two $\bd\to\pi^+\pi^-$ CP
asymmetries ${\cal A}_{dir}$ and ${\cal A}_{mix}$ depend on only two
quantities. By measuring these three observables, one can test for the
presence of physics beyond the standard model. In this paper, using
present data, and including theoretical uncertainties due to SU(3)
breaking, we perform such an analysis. The experimental errors are
still very large, but with improved measurements from Babar, Belle and
CDF this will be a useful method to search for new physics.
\end{abstract}

\maketitle

At present, we do not have reliable theoretical computations of
individual $B$ decay rates. For example, calculations within QCD
factorization \cite{BBNS} predict $BR(B^0 \to \pi^+ \pi^-)$ to be
larger, and $BR(B^0 \to \pi^0 \pi^0)$ smaller, than what is found
experimentally \cite{BN}. This may imply the presence of physics
beyond the standard model (SM), or there might be a problem with QCD
factorization. Indeed, in Ref.~\cite{BFRS}, it is argued that new
physics is unnecessary. Rather, large nonfactorizable effects can
account for the experimental branching ratios. On the other hand,
calculations within soft-collinear effective theory yield $B\to\pi\pi$
branching ratios which are consistent with the data \cite{BDRS}
without invoking nonfactorizable effects of the type discussed in
Ref.~\cite{BFRS}.

This brief discussion underscores the fact that the calculation of
rates for individual $B$ decays is very difficult.  If one wishes to
search for new physics, it is better to look at ratios of decay
rates. If, due to some symmetry, the ratio of two rates is exactly
predicted in the SM, then any deviation from this prediction would be
a clear signal of physics beyond the SM.

It is therefore natural to try to relate $B^0 \to \pi^+\pi^-$ to
another decay using flavor SU(3) symmetry. The obvious partner decay
is $B^0 \to K^+ \pi^-$ \cite{FleischerBKpi}. In terms of diagrams, the
amplitudes for these two processes are \cite{GHLR}
\bea
A(\bd \to \pi^+ \pi^-) & = & - T - P - E - PA - {2 \over 3} \pewc ~,
\nn\\
A(\bd \to K^+ \pi^-) & = & - T' - P' - {2 \over 3} \pewcp ~.
\label{amps}
\eea
In the above, the amplitudes are written in terms of a color-favored
tree amplitude $T$, a gluonic penguin amplitude $P$, an exchange
amplitude $E$, a penguin annihilation amplitude $PA$, and a
color-suppressed electroweak penguin amplitude $\pewc$. For ${\bar b}
\to {\bar d} u {\bar u}$ ($B^0 \to \pi^+ \pi^-$) and ${\bar b} \to
{\bar s} u {\bar u}$ ($B^0 \to K^+ \pi^-$), we write the diagrams with
no primes and primes, respectively. If one assumes that the spectator
quark plays no role in the decay, then the nonfactorizable
contributions $E$ and $PA$ are negligible. In this case, the various
contributions to the two amplitudes are equal, apart from elements of
the Cabibbo-Kobayashi-Maskawa (CKM) matrix. However, the neglect of
$E$ and $PA$ is problematic. If nonfactorizable effects do indeed play
an important role in $B^0 \to \pi^+\pi^-$ \cite{BFRS}, then the SU(3)
relations between the $B^0 \to \pi^+\pi^-$ and $B^0 \to K^+ \pi^-$
amplitudes are invalidated.

Naive estimates of $P$ and $PA$ put them at $\lsim 5$\% of the
dominant contributions \cite{GHLR}. However, their size is really an
experimental question. Some decays (e.g.\ $\bd\to D_s^+ D_s^-$, $\bd
\to K^+ K^-$, $\bs \to \pi^+\pi^-$, etc.) proceed mainly through
exchange-type interactions involving the spectator quark. The
measurement of these rates will provide an estimate of the size of
exchange-type contributions.

Because of the uncertainty about nonfactorizable effects, a better
partner decay is $\bs \to K^+ K^-$. Its amplitude is
\beq
A(\bs \to K^+ K^-) = - T' - P' - E' - PA' - {2 \over 3} \pewcp ~.
\eeq
Thus, even including nonfactorizable effects, the amplitude for this
decay is related by SU(3) (U-spin) to that for $\bd \to \pi^+\pi^-$
\cite{Uspin}. It is therefore possible to relate observables measured
in one decay to those in the other decay. The only theoretical
uncertainty is the breaking of SU(3) \cite{Uspin,SU3break}.

The idea of relating the amplitudes of $\bd \to \pi^+\pi^-$ and $\bs
\to K^+ K^-$ via SU(3) is not new. It has been used in
Refs.~\cite{BFRS,Uspin,FM0,FM,BsKK} to relate observables in the two
decays. In particular, the CP asymmetries in $\bs(t) \to K^+ K^-$ can
be written in terms of the same parameters that appear in the CP
asymmetries for $\bd(t) \to \pi^+\pi^-$ (up to SU(3)-breaking effects
\cite{Uspin,SU3break}). In the present paper we note that the
branching ratio (BR) for $\bs\to K^+ K^-$ can be written in terms of
the observables found in $\bd(t) \to \pi^+\pi^-$. Thus, the
measurements of the BR and CP asymmetries in $\bd(t) \to \pi^+\pi^-$
and the BR for $\bs\to K^+ K^-$ provide a straightforward consistency
check of the SM. Furthermore, all four of these quantities have
already been measured so that it is possible to apply this test
now. (Unfortunately, as we will see, because of the large experimental
errors, no firm conclusions can be drawn at present.)

Within the SM, the decay $\bd\to \pi^+\pi^-$ receives charged-current
contributions, proportional to $V_{ub}^* V_{ud}$, and penguin
contributions $V_{ub}^* V_{ud} P_u + V_{cb}^* V_{cd} P_c + V_{tb}^*
V_{td} P_t$. (The charged-current term includes the $T$ and $E$
diagrams of Eq.~(\ref{amps}). Similarly, the penguin term includes
$P$, $PA$ and $\pewc$.) Using CKM unitarity to eliminate $P_t$, the
penguin contributions become $V_{ub}^* V_{ud} (P_u - P_t) + V_{cb}^*
V_{cd} (P_c - P_t)$. The amplitude for $\bd\to \pi^+\pi^-$ can then
be written \cite{Uspin}
\bea
A(\bd\to \pi^+\pi^-) & = & V_{ub}^* V_{ud} (A_{\rm CC}^{u} +
A_{\rm pen}^{ut}) + V_{cb}^* V_{cd} A_{\rm pen}^{ct} \nn\\
& = & {\cal C} \left(e^{i \gamma} - d e^{i\theta} \right) ~,
\label{ampBpipi}
\eea
where $A_{\rm pen}^{it} \equiv P_i - P_t$ ($i=u,c$), $\gamma$ is a CP
phase \cite{pdg}, and
\bea
{\cal C} & \equiv & |V_{ub}^* V_{ud}| \left( A_{\rm CC}^{u} + A_{\rm
pen}^{ut} \right) ~, \nn\\
d e^{i\theta} & \equiv & \frac{1}{R_b} \left( \frac{A_{\rm pen}^{ct}}
{A_{\rm CC}^u + A_{\rm pen}^{ut}} \right) ~,
\eea
with $R_b = |(V_{ub}^* V_{ud}) / (V_{cb}^* V_{cd}) |$. In the above,
$\theta$ is the relative strong phase between $A_{\rm pen}^{ct}$ and
$(A_{\rm CC}^u + A_{\rm pen}^{ut})$, and ${\cal C}$ contains a strong
phase. The amplitude for $\bdbar \to \pi^+ \pi^-$ can be obtained from
Eq.~(\ref{ampBpipi}) by changing the sign of $\gamma$:
\beq
A(\bdbar\to \pi^+\pi^-) = {\cal C} \left(e^{-i \gamma}
- d e^{i\theta} \right) ~.
\label{ampBbarpipi}
\eeq

There are both direct and mixing-induced CP asymmetries in $\bd(t) \to
\pi^+ \pi^-$. Defining
\bea
{\cal A}^{+-}(t) & = & { \Gamma(\bd(t) \to \pi^+\pi^-) -
\Gamma(\bdbar(t) \to \pi^+\pi^-) \over \Gamma(\bd(t) \to \pi^+\pi^-) +
\Gamma(\bdbar(t) \to \pi^+\pi^-) } \nn\\
& = & {\cal A}_{dir} \cos \Delta M t + {\cal A}_{mix} \sin \Delta M t
~,
\eea
we have
\bea
{\cal A}_{dir} & \!=\! & -{2 d \sin \theta \sin \gamma \over 1 - 2 d
\cos\theta \cos\gamma + d^2} ~, \\
{\cal A}_{mix} & \!=\! & { \sin(\phi_d + 2 \gamma) -2 d
\cos\theta \sin(\phi_d + \gamma) + d^2 \sin \phi_d \over 1 - 2 d
\cos\theta \cos\gamma + d^2} ~, \nn
\eea
where $\phi_d$ is the phase of $\bd$--$\bdbar$ mixing.

The decay $\bd\to\pi^+\pi^-$ is related to $\bs\to K^+ K^-$ by flavor
SU(3). That is, we can write \cite{Uspin}
\beq
A(\bs\to K^+ K^-) = \left\vert {V_{us} \over V_{ud}} \right\vert
{\cal C}' \left(e^{i \gamma} +
\left\vert {V_{cs} V_{ud} \over V_{us} V_{cd}} \right\vert
d' e^{i\theta'} \right) ~,
\eeq
where
\bea
{\cal C}' & \equiv & |V_{ub}^* V_{ud}| \left( A_{\rm CC}^{\prime u} +
A_{\rm pen}^{\prime ut} \right) ~, \nn\\
d' e^{i\theta'} & \equiv & \frac{1}{R_b} \left( \frac{A_{\rm
pen}^{\prime ct}} {A_{\rm CC}^{\prime u} + A_{\rm pen}^{\prime ut}}
\right) ~.
\eea
In the SU(3) limit, we have
\beq
{\cal C} = {\cal C'} ~,~~ d e^{i\theta} = d' e^{i\theta'} ~.
\eeq
Thus, observables in the decay $\bs\to K^+ K^-$ are related to
observables in $B^0 \to \pi^+ \pi^-$. (Of course, SU(3)-breaking
effects must also be taken into account. We do this below.)

We now define the ratio of BR's for $\bs\to K^+ K^-$ and
$\bd\to\pi^+\pi^-$:
\beq
R_d^s \equiv {\langle BR(B_s \to K^+ K^-) \rangle \over \langle BR(B_d
\to \pi^+ \pi^-) \rangle} ~,
\eeq
where the symbol $\langle ... \rangle$ indicates an average over $B^0$
and ${\bar B}^0$ decays. In terms of theoretical parameters, this
ratio is given by
\beq
R^s_d = {1 \over \epsilon} \left\vert {{\cal C}' \over 
{\cal 
C}}
  \right\vert^2 { \epsilon^2 + 2 \epsilon d' \cos\theta' 
\cos\gamma +
 {d'}^2 \over 1 - 2 d \cos\theta \cos\gamma + d^2 } 
f_{\sss PS} ~.
\eeq
where $\epsilon=\lambda^2/(1-\lambda^2)$ and
$\lambda=0.22$ is the Cabibbo
angle. The quantity 
\beq 
f_{\sss PS}={M_{Bd}\over M_{Bs}} \times {\tau_{Bs}\over \tau_{Bd}}
\times {\phi(M_K/M_{Bs},M_K/M_{Bs})\over
\phi(M_\pi/M_{Bd},M_\pi/M_{Bd})}
\eeq 
is a (known) phase-space factor (see \cite{Uspin,FM}), which takes
value $f_{\sss PS} = 0.92$. (The experimental quantity $R_d^s$ is
related to the theoretical quantity $H$ of Ref.~\cite{Uspin}.)

In the SU(3) limit, the three observables ${\cal A}_{dir}$, ${\cal
A}_{mix}$ and $R^s_d$ all depend on the four theoretical parameters
$d$, $\theta$, $\phi_d$ and $\gamma$. However, $\phi_d$ has been
measured through CP violation in $\bd(t) \to J/\psi \ks$ and, assuming
the SM, we can take the value of $\gamma$ from independent
measurements. Then knowledge of two measurements allows us to {\it
predict} the third. We can then compare this prediction with the
experimental measurement to see if it is consistent. (Equivalently, we
can solve for $d$, $\theta$ and $\gamma$ using the three
observables. We can then check for consistency by comparing the value
of $\gamma$ with that obtained elsewhere.)

Note that, if an inconsistency is found, this analysis does not
pinpoint the origin of the physics beyond the SM. There are several
places in which this new physics could enter: the extractions of
$\phi_d$ or $\gamma$, or the amplitudes for $B^0\to\pi\pi$ or $\bs\to
K^+ K^-$. If a discrepancy is found, further experimental tests will
be necessary to identify the new physics.

It is possible to explicitly write $R^s_d$ in terms of ${\cal
A}_{dir}$ and ${\cal A}_{mix}$ \cite{FM}, but the analytic expression
is not terribly illuminating. It is better to use figures to give the
reader a sense of what the relation between ${\cal A}_{dir}$, ${\cal
A}_{mix}$ and $R^s_d$ looks like.

We first assume perfect SU(3) symmetry, with one exception. The ratio
$\vert {\cal C}'/{\cal C} \vert$ has recently been calculated using
QCD sum rules \cite{Mannel}:
\beq
\left\vert {{\cal C}' \over {\cal C}} \right\vert =
1.76^{+0.15}_{-0.17} ~.
\label{CprimeCratio}
\eeq
We take $\vert {\cal C}'/{\cal C} \vert = 1.76$ (with no error). The
measurement of CP violation in $\bd(t) \to J/\psi \ks$ gives $\sin
\phi_d = 0.73 \pm 0.05$ \cite{pdg}. We take $\phi_d = 2\beta =
47^\circ$. (If $\phi_d = 133^\circ$ is taken, this would already be
evidence for physics beyond the SM \cite{FIM}.) We also assume $\gamma
= 65^\circ$, which is the value preferred by independent measurements
($\epsilon_K$, $V_{cb}$, $|V_{ub}/V_{cb}|$, etc.) \cite{gamma}.

In Fig.~1a we present the (correlated) allowed values of $R^s_d$ and
${\cal A}_{dir}$, for various values of ${\cal A}_{mix}$. Fig.~1b is
similar, but with ${\cal A}_{dir} \leftrightarrow {\cal A}_{mix}$. In
both cases, the values of the CP asymmetries respect \cite{GroRosner}
\beq
{\cal A}_{dir}^2 + {\cal A}_{mix}^2 \le 1 ~.
\label{CPconstraints}
\eeq
{}From these figures, one sees that the equation relating ${\cal
A}_{dir}$, ${\cal A}_{mix}$ and $R^s_d$ is independent of the sign of
${\cal A}_{dir}$, but not of ${\cal A}_{mix}$. Note also that this is
a bi-value relation: for given values of ${\cal A}_{dir}$ and ${\cal
A}_{mix}$, there are two possibilities for $R^s_d$. This will be
important when we add errors below. Finally, in Fig.~1c, we show the
(correlated) allowed values of ${\cal A}_{dir}$ and ${\cal A}_{mix}$,
for various values of $R^s_d$.

\begin{figure}[t]
\vskip1.4truecm
\epsfxsize 3.0truein \epsfbox[84 572 329 732]{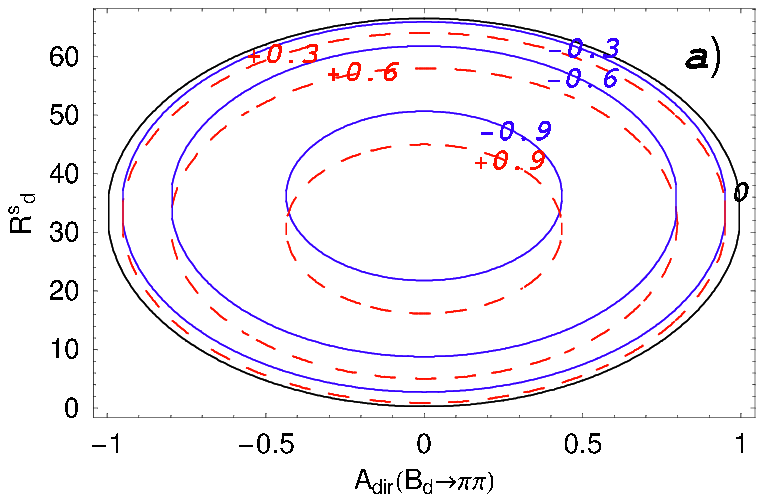}
\epsfxsize 3.0truein \epsfbox[84 572 329 732]{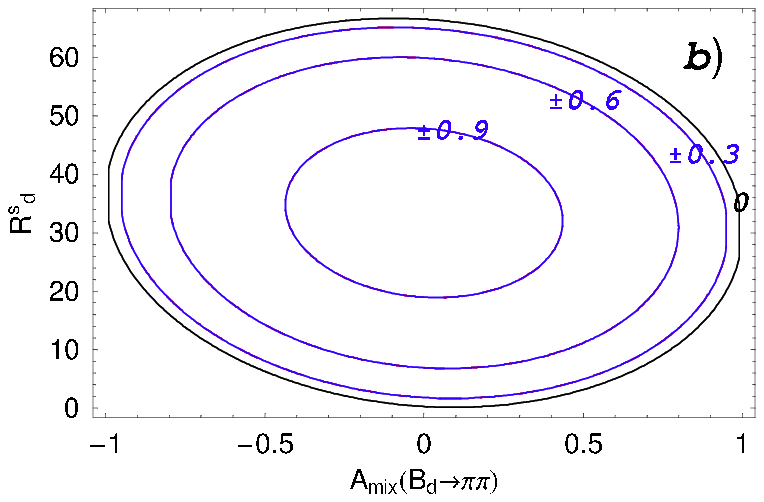}
\epsfxsize 3.0truein \epsfbox[84 572 329 732]{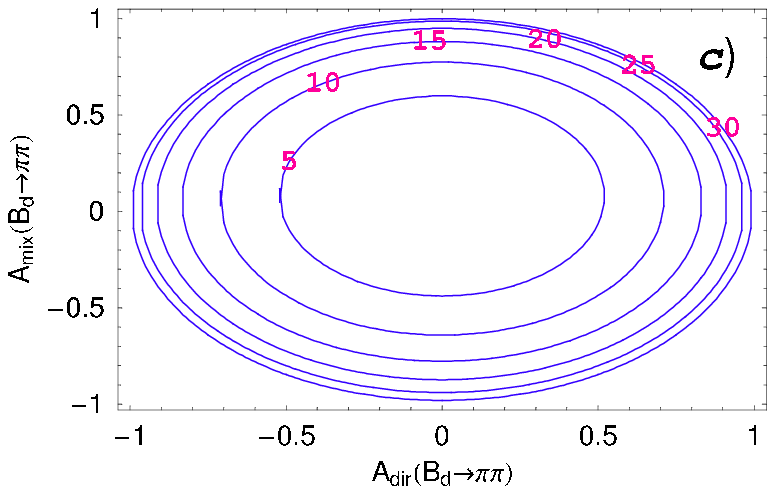}
\vskip-2.0truecm
\caption{The (correlated) allowed values of a) $R^s_d$ and ${\cal
A}_{dir}$, b) $R^s_d$ and ${\cal A}_{mix}$, c) ${\cal A}_{dir}$ and
${\cal A}_{mix}$ in the SU(3) limit (except for $\vert {\cal C}'/{\cal
C} \vert$). The values assumed for the third observable, a) ${\cal
A}_{mix}$, b) ${\cal A}_{dir}$ and c) $R^s_d$, are shown on the
figures.}
\end{figure}

Of course, there are errors on the input value of $\gamma$. We take
$\gamma = (65 \pm 7)^\circ$ \cite{gamma}. (However, we still assume
that $2\beta = 47^\circ$, with no error.) We also include the error on
the SU(3)-breaking quantity $\vert {\cal C}'/{\cal C} \vert$
[Eq.~(\ref{CprimeCratio})]. There is also SU(3) breaking in comparing
$d$ and $d'$, and $\theta$ and $\theta'$. We assume $d'/d = 1.0 \pm
0.2$, and $\theta' - \theta = 0^\circ \pm 40^\circ$ \cite{footnote}.
Normally, theoretical errors imply that the true value lies somewhere
in the ``$1\sigma$'' range. Thus, in our numerical analysis, we allow
the various quantities to take any values in the following ranges:
$58^\circ \le \gamma \le 72^\circ$, $1.49 \le {\cal C}' / {\cal C} \le
1.91$, $0.8 \le d'/d \le 1.2$, $-40^\circ \le \theta' - \theta \le
40^\circ$.

When these uncertainties are included, the ${\cal
A}_{dir}$--${\cal A}_{mix}$--$R^s_d$ relation shown in Figs.~1a-1c
gets smeared out. In Fig.~2, we present the allowed
$R^s_d$--${\cal A}_{dir}$ region for three values of ${\cal
A}_{mix}$: 0.1 (area inside the dashed line), 0.5 (dotted line)
and 0.9 (solid line). For simplicity we consider only ${\cal
A}_{dir} < 0$, so this plot can be thought of as the left half of
Fig.~1a for the case where uncertainties are included in the
analysis. From this plot, we see that, due to the theoretical
error, the curves of Fig.~1a become wide regions. (For ${\cal
A}_{mix} = 0.1$ and 0.5, there is still an area of $R^s_d$--${\cal
A}_{dir}$ parameter space in the center of the region which is not
allowed; for ${\cal A}_{mix} = 0.9$, there is no such area.)

\begin{figure}[t]
\vskip1.4truecm
\epsfxsize 3.0truein \epsfbox[84 572 329 732]{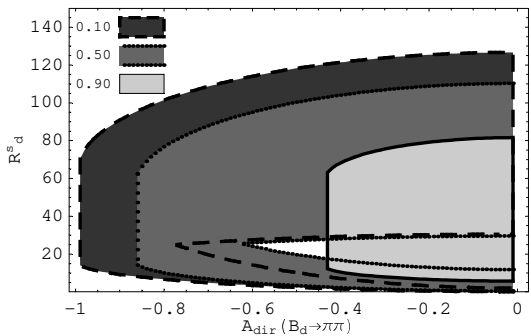}
\vskip-2.0truecm
\caption{The (correlated) allowed values of $R^s_d$ and ${\cal
A}_{dir}$ for three values of ${\cal A}_{mix}$. We take the
experimental and theoretical uncertainties into account: $58^\circ \le
\gamma \le 72^\circ$, $1.49 \le {\cal C}' / {\cal C} \le 1.91$, $0.8
\le d'/d \le 1.2$, $-40^\circ \le \theta' - \theta \le 40^\circ$. For
${\cal A}_{mix} = 0.1$, 0.5 and 0.9, the allowed $R^s_d$--${\cal
A}_{dir}$ regions are, respectively, inside the dashed, dotted and
solid lines.}
\end{figure}

Up to now, the discussion has been purely theoretical. However,
experimental data is presently available for ${\cal A}_{dir}$, ${\cal
A}_{mix}$ and $R^s_d$. For the $\bd\to\pi^+\pi^-$ CP asymmetries, we
have
\bea
{\cal A}_{dir} & = & \cases{
-0.19 \pm 0.19 \pm 0.05 & BaBar \cite{BaBarCP}, \cr
-0.58 \pm 0.15 \pm 0.07 & Belle \cite{BelleCP}. \cr} \nn\\
{\cal A}_{mix} & = & \cases{
+0.40 \pm 0.22 \pm 0.03 & BaBar \cite{BaBarCP}, \cr
+1.00 \pm 0.21 \pm 0.07 & Belle \cite{BelleCP}. \cr}
\label{CPasymexp}
\eea
For the branching ratios, CDF recently reported \cite{CDF}
\beq
\left( {f_d \over f_s} \right) {BR(B^0 \to \pi^+\pi^-) \over BR(\bs \to
K^+ K^-)} = 0.35 \pm 0.18 ~.
\eeq
Taking $f_s/f_d = 0.27 \pm 0.04$ \cite{pdg}, we find
\beq
R_d^s = 10.6 \pm 5.7 ~.
\label{Rdsexp}
\eeq
We can now compare the value of $R_d^s$ implied by the measurements of
the CP asymmetries [Eq.~(\ref{CPasymexp})] with that measured directly
[Eq.~(\ref{Rdsexp})]. A discrepancy would signal the presence of new
physics.

The current situation is shown in Fig.~3. We treat separately the
Belle and BaBar measurements of the CP asymmetries
[Eq.(\ref{CPasymexp})]. We allow each of ${\cal A}_{dir}$ and ${\cal
A}_{mix}$ to take values in their $\pm 1\sigma$ ranges. Including all
theoretical uncertainties, this yields a range of allowed values for
$R_d^s$. Note that the BaBar region is quite a bit bigger than that
of Belle. This is because a large part of the Belle ${\cal
A}_{dir}$--${\cal A}_{mix}$ parameter space is already ruled out,
having violated the constraint of Eq.~(\ref{CPconstraints}).
Superimposed on this is the CDF result for $R_d^s$
[Eq.(\ref{Rdsexp})], also given for $\pm 1\sigma$. (We also show the
case where the SU(3)-breaking effect is smaller than expected, $0.9
\le d'/d \le 1.1$, but this does not change things much.)

\begin{figure}[t]
\vskip1.4truecm
\epsfxsize 3.0truein \epsfbox[84 572 329 732]{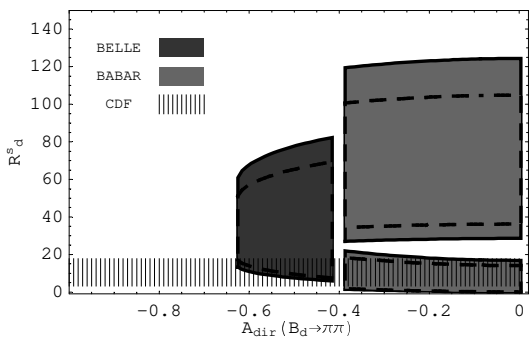}
\vskip-2.0truecm
\caption{The allowed values of $R^s_d$, including all theoretical
uncertainties, as derived from measurements of ${\cal A}_{dir}$ and
${\cal A}_{mix}$ [Eq.(\protect\ref{CPasymexp})]. The Belle and BaBar
data are treated separately, and the CP asymmetries are allowed to
vary within their $\pm 1\sigma$ ranges. The CDF measurement of $R^s_d$
[Eq.(\protect\ref{Rdsexp})] is superimposed (straight lines). The
internal dashed line corresponds to the case where one has smaller
SU(3) breaking, $0.9 \le d'/d \le 1.1$.}
\end{figure}

It is clear from this figure that, at present, there is no discrepancy
-- the measured value of $R^s_d$ overlaps with the BaBar and Belle
regions. That is, the values of $R^s_d$ predicted by measurements of
the CP asymmetries include those of Eq.~(\ref{Rdsexp}). On the other
hand, the experimental errors are still very large. As these errors
get reduced, this analysis will become increasingly useful as a method
of searching for physics beyond the SM.

\smallskip
We thank Giovanni Punzi for helpful discussions. JM thanks C. Burgess
for the hospitality of McGill University, where part of this work was
done. This work was financially supported by NSERC of Canada (DL) and
by FPA2002-00748 and the Ramon y Cajal Program (JM).


\begin{thebibliography}{References:}
\bibitem{footnote} The assumed size of SU(3) breaking in $d'/d$ and
$\theta' - \theta$ is smaller than that found for $|{\cal C}'/{\cal
C}|$ in Eq.~(\ref{CprimeCratio}). This is because all factorizable
SU(3)-breaking effects cancel in the ratio $d'/d$, leaving only
nonfactorizable corrections. SU(3) breaking in $|{\cal C}'/{\cal C}|$
is expected to be larger since both factorizable and nonfactorizable
terms are present.
\end{thebibliography}
\end{document}